# Opening Scholarly Communication in Social Sciences: Supporting Open Peer Review with Fidus Writer

Philipp Mayr[1], Fakhri Momeni, Christoph Lange

## Introduction

Scholarly communication in the social sciences is centered around publications, in which data also play a key role. The increasingly collaborative scientific process, from a project plan, to collecting data, to interpreting them in a paper and submitting it for peer review, to publishing an article, to, finally, its consumption by readers, is insufficiently supported by contemporary information systems. They support every individual step, but media discontinuities between steps cause inefficiency and loss of information: word processors lack direct access to data; reviewers cannot provide feedback inside the environment in which authors revising their papers; open access web publishing is constrained to document formats designed for paper printing but neglecting the Web's accessibility and interactivity potential; finally, readers, seeing a single frozen view of the underlying data in a paper, are unable to access the full extent of the data and to make observations beyond the restricted scope chosen by the author.

## The OSCOSS project

Web technology can address these problems. Isolated solutions, such as tools for publishing data on the Web for easy retrieval and visualization, exist in preliminary manifestations in the social sciences, but have not been integrated into tools for writing, reviewing and publishing articles. Tools that assist writers in making their documents' structure explicit for information systems, as well as document browsers that use articles as interactive interfaces to related information on the Web have been successfully deployed in the life sciences. In the Opening Scholarly Communication in Social Sciences (OSCOSS) project[2], funded by DFG, we will transfer these ideas to the social sciences (see Garcia et al., 2012) by integrating existing data and publication management services into a web-based collaborative writing environment that publishers can set up to supports all types of end users throughout the publication process: authors, reviewers and readers. The OSCOSS project aims at providing integrated support for all following steps:

1. collaborative writing of a scientific paper,

2. collecting data related to existing publications,

3. interpreting and including data in a paper,

4. submitting the paper for peer review,

5. reviewing the paper,

---

[1] Corresponding author: philipp.mayr@gesis.org

[2] http://www.gesis.org/en/research/external-funding-projects/overview-external-funding-projects/oscoss/

6. publishing an article, and, finally,

7. facilitating its consumption by readers.

With the collaborative document editor Fidus Writer[3] (see Wilm and Frebel, 2015; Perkel, 2014) and Open Journal Systems[4] system we choose a stable technical foundation (see Figure 1). We secure user acceptance by respecting the characteristics of the traditional processes social scientists are used to: web publications must have the same high-quality layout as print publications, and information must remain citable by stable page numbers. To ensure we meet these requirements, we will work closely with the publishers of "methods, data, analyses" (mda)[5] and "Historical Social Research" (HSR)[6], two international peer reviewed open access journals published by GESIS, and build early demonstrators for usability evaluation.

Our system will initially provide readers, authors and reviewers with an alternative, thus having the potential to gain wider acceptance and gradually replace the old, incoherent publication process of our journals and of others in related fields. It will make journals more "open" (in terms of reusability) that are open access already, and it has the potential to serve as an incentive for turning "closed" journals into open access ones.

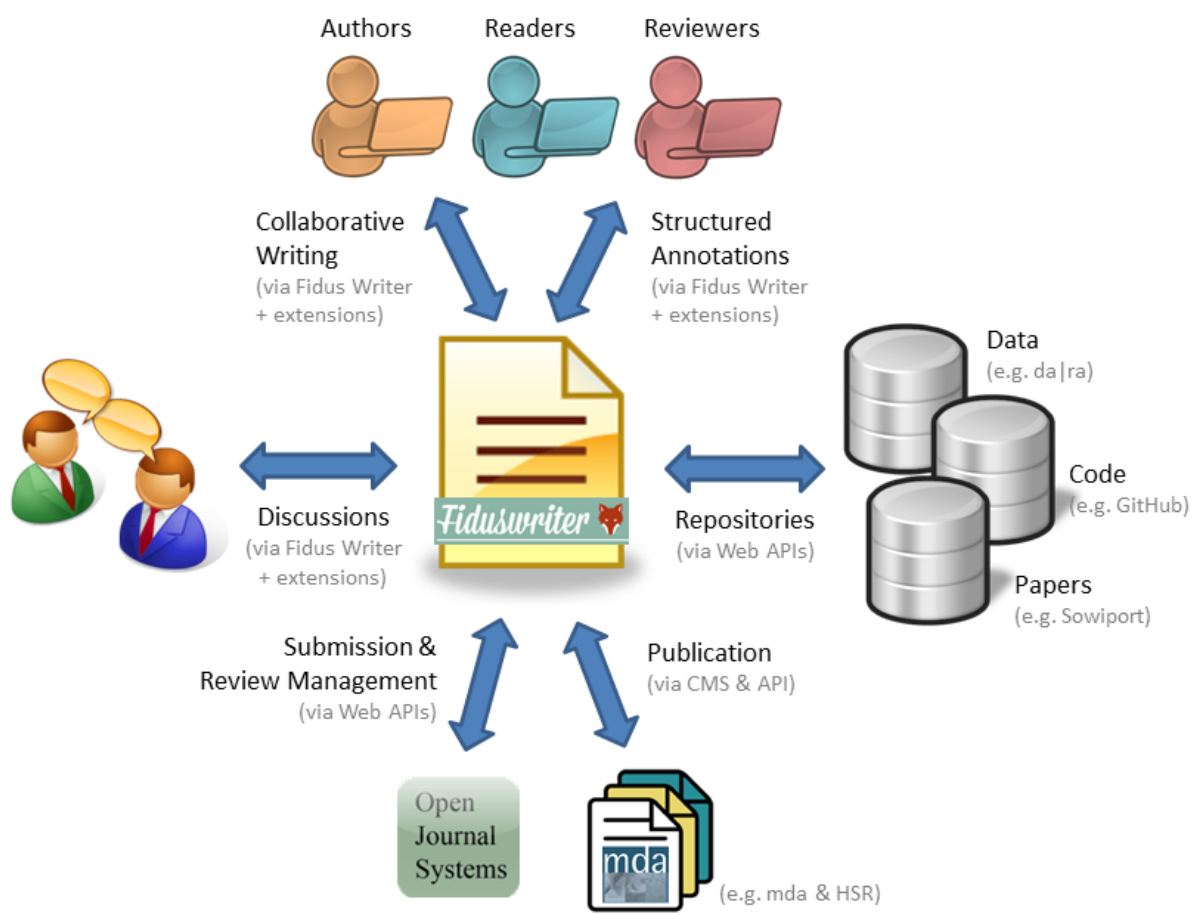

**Figure 1**: System architecture

---

[3] http://fiduswriter.org
[4] https://pkp.sfu.ca/ojs/
[5] http://www.gesis.org/publikationen/zeitschriften/mda/
[6] http://www.gesis.org/hsr/hsr-home/

# Use Cases

To justify the relevance of our work and to explain in what ways our proposed project opens scholarly communication and publishing from multiple perspectives, we first analyze one real world example, and then present three idealized use cases from the perspective of three major stakeholders of scholarly communication and publishing: reader, author and reviewer.

## Real world example

In a recent article in the mda journal, Kroll (2011) cites a dataset by name, writing "Der Index wird zuletzt auf Basis der Daten der BIBB/BAuA-Erwerbstätigenbefragung 2006 und des Telefonischen Gesundheitssurveys „Gesundheit in Deutschland Aktuell" (GEDA) 2009 des Robert Koch-Instituts anhand von Gesundheitsindikatoren intern und extern validiert." and he cites other articles citing the same dataset ("zu den Datensätzen vgl. Hall 2009; Kurth et al. 2009; RKI 2010"; Kroll 2001, p.67). But an explicit link to the dataset is missing in his article. Still, the dataset is registered in da|ra[7], a registry for raw datasets hosted by GESIS, and can be looked up; it has a metadata record and a DOI. In the concrete case the dataset can be analyzed after signing a contract.

As one of the first results of the OSCOSS project, we have implemented a semi-automated link detection approach for detecting references to datasets in paper full texts, and for linking them to the right datasets in da|ra (Ghavimi et al. 2016).

## Reader Use Case

Mark is conducting a survey on "paradata"[8]. He has selected a number of relevant articles that have been published recently in mda and is now studying them in detail. He wants to focus on observations with a high statistical significance. In one article, he has identified an interesting "non-response bias", which is presented in a table. For his future work, he wants to precisely bookmark this occurrence as "useful for my survey", and add an annotation that helps him remember what exactly was interesting and why.

## Author Use Case

Jakob has a draft of a paper, written in Word, and wants to extend it by performing a different analysis on those base data that Arthur, a researcher from the same community, has used for an earlier publication. Arthur, in his publication, cites a dataset from da|ra. The dataset has a DOI, and furthermore, Arthur describes in his paper what chunks of the dataset he based his analysis on, and what analysis method he applied. The R code that implements the data analysis is open source and available from an online repository. Arthur presented the output of this analysis as a table and a diagram in his paper. Jakob invites his co-author Dagmar to do a different analysis of the same data: Dagmar re-applies the same analysis method but changes the values of some regression parameters, and Jakob compares the result of this analysis to the result of Arthur's analysis. Their new article includes a new table, and a copy of Arthur's one, side by side and citing Arthur's original table and the underlying dataset, and draws new conclusions. They submit their article to mda.

---

[7] http://www.da-ra.de/en/home/
[8] The paradata of a survey are data about the process by which the survey data were collected. Paradata are usually "administrative data about the survey". (http://en.wikipedia.org/wiki/Paradata)

## Reviewer Use Case

Jakob and Dagmar submitted their manuscript, pointing to data and the R code, to the journal mda. Rainer gets assigned Jakob's and Dagmars's manuscript for review. He wants to check whether Jakob and Dagmar have done their analysis in a correct way. He downloads their R code and raw data and redoes the calculation described in the paper. He observes that, for one of the statements that Jakob and Dagmar have made in the results section of their manuscript, the R output of the analysis does not give sufficient evidence. They should have known from Ariane's paper published two years ago, that in one of the items of the dataset that they analyzed, some data items are too sparse for reliably applying the significance test to them. Rainer marks the respective statement in Jakob's and Dagmar's result section, adding a reference to the methodology section of Ariane's paper and to the affected item in the dataset. Finally, the editorial board decides to accept the submission, given that a major revision is made. Jakob and Dagmar receive the paper with 100 comments attached. As the comments are attached to precise parts of the paper, grouped by reviewers and classified as "major" vs. "minor", they can quickly prioritize the necessary tasks to improve their article.

In this poster we will present the framework of the system and highlight the reviewer use case. In addition, we will be ready to give a live demo of relevant features of the systems on which the OSCOSS collaboration environment will be based.

## Acknowledgement

This work was funded by DFG, grant no. SU 647/19-1 and AU 340/9-1; the OSCOSS project at GESIS and University Bonn.

## References

Garcia, A., Mayr, P., & Garcia, L. J. (2012). Simple Semantic Enrichment of Scientific Papers in Social Sciences. In SWIB 2012. Cologne, Germany. Retrieved from http://swib.org/swib12/

Ghavimi, B., Mayr, P., Vahdati, S., & Lange, C. (2016). Identifying and Improving Dataset References in Social Sciences Full Texts. Pre-print, doi:10.5281/zenodo.44608

Kroll, L. E. (2011). Konstruktion und Validierung eines allgemeinen Index für die Arbeitsbelastung in beruflichen Tätigkeiten auf Basis von ISCO-88 und KldB-92. Methoden, Daten, Analysen (mda), 5(1), 63–90. Retrieved from http://nbn-resolving.de/urn:nbn:de:0168-ssoar-255027

Perkel, J. M. (2014). Scientific writing: the online cooperative. Nature, 514(7520), 127–128. doi:10.1038/514127a

Wilm, J., & Frebel, D. (2015). Real-world challenges to collaborative text creation. In Proceedings of the 2nd International Workshop on (Document) Changes: modeling, detection, storage and visualization - DChanges '14 (pp. 1–4). New York, New York, USA: ACM Press. doi:10.1145/2723147.2723154